\documentclass[aps,final,twocolumn,10pt,superscriptaddress,citeautoscript,flushbottom]{revtex4-1}%

\usepackage{amssymb,amsmath,amsfonts}
\usepackage{graphicx,color}
\usepackage{tabularx}

\newcommand{\1}{\begin{equation}}
\newcommand{\2}{\end{equation}}
\newcommand{\ea}{\begin{eqnarray}} 
\newcommand{\ee}{\end{eqnarray}}

\renewcommand{\vec}[1]{\mathbf{#1}}
\newcommand{\abs}[1]{\left| #1 \right|} % for absolute value
\newcommand{\avg}[1]{\left< #1 \right>} % for average
\let\baraccent=\= % rename builtin command \= to \baraccent
\renewcommand{\=}[1]{\stackrel{#1}{=}} % for putting numbers above =
\usepackage{color}
 \definecolor{blue}{rgb}{0,0,1} %%{\color{blue} text...}
 \definecolor{sepia}{rgb}{0,0.8,0.2}
 \definecolor{redi}{rgb}{0.5176,0.0078,0.0078}

\begin{document} 
\title{Taming Polar Active Matter with Moving Substrates: Directed Transport and Counterpropagating Macrobands}

\author{Alexandra Zampetaki} 
\affiliation{Zentrum  f\"{u}r  Optische  Quantentechnologien,  Universit\"at Hamburg, Luruper  Chaussee  149,  22761  Hamburg,  Germany}
%\email{alexandra.zampetaki@physnet.uni-hamburg.de}
\author{Peter Schmelcher} 
\affiliation{Zentrum  f\"{u}r  Optische  Quantentechnologien,  Universit\"at Hamburg, Luruper  Chaussee  149,  22761  Hamburg,  Germany}
\author{Hartmut L\"owen}
\affiliation{Institut f\"{u}r Theoretische Physik II: Weiche Materie, Heinrich-Heine-Universit\"{a}t D\"{u}sseldorf, D-40225 D\"{u}sseldorf, Germany}
\author{Benno Liebchen}
\affiliation{Institut f\"{u}r Theoretische Physik II: Weiche Materie, Heinrich-Heine-Universit\"{a}t D\"{u}sseldorf, D-40225 D\"{u}sseldorf, Germany}
%\email{liebchen@hhu.de}

\date{\today}

\begin{abstract}
%The quest for 
%the control of the individual dynamics of active particles with external fields is among the most explored topics in active matter. 
Following the goal 
of using active particles as targeted cargo carriers aimed, for example, to deliver drugs towards cancer cells, the quest for 
the control of individual active particles with external fields is among the most explored topics in active matter. 
Here, we provide a scheme allowing to control collective behaviour in active matter, focusing on the fluctuating band patterns 
naturally occurring e.g. in the Vicsek model. %active polar particles.  
We show that exposing these patterns to a travelling wave potential tames them, yet in a remarkably nontrivial way: 
the bands, which initially pin to the potential and comove with it, upon subsequent collisions, self-organize into a macroband, 
featuring a predictable transport against the direction of motion of the travelling potential. 
Our results provide a route to simultaneously control transport and structure, i.e. micro- versus macrophase separation, in polar active matter.
\end{abstract}

\maketitle

\section{Introduction}
Active matter contains self-propelled particles like bacteria, algae, or synthetic autophoretic Janus colloids whose properties can be designed
on demand
\cite{Ramaswamy2010,Marchetti2013,Bechinger2016}. 
As one of their main characteristics, these systems are intrinsically out of equilibrium 
allowing them to self-organize into new
ordered and even functional structures. %, the latter being frequently encountered in the biological world. 
In synthetic active systems, such structures
include dynamic clusters which dynamically form and break-up in low density Janus colloids
\cite{Theurkauff2012,Palacci2013,Buttinoni2013,Ginot2018,Liebchen2018} 
as well as laser driven colloids which spontaneously start 
to move ballistically (self-propel) when binding together \cite{Soto2014,Varma2018,Schmidt2018}.
Likewise, biological microswimmers form patterns such as vortices in
bacterial turbulence \cite{Sokolov2012,Wensink2012,Kaiser2014,Stenhammar2017}, 
or swirls and microflock patterns in chiral active matter like curved polymers or sperm \cite{Riedel2005,Loose2014,Denk2016,LiebchenLevis2017}. 

Much of what we know about active systems and the patterns they form roots in explorations of minimal models which to some extend represent broader classes of active systems 
showing the same symmetries.
The pioneering example of such a minimal model is the Vicsek model describing 
polar self-propelled particles such as actin-fibres mixed with motor proteins \cite{Schaller2010,Goff2016}, certain microorganisms \cite{Toner2005},
self-propelled rods \cite{Paxton2004,Peruani2016} or ``birds'' \cite{Toner1995,Toner2005} 
which only see their neighbors and have a tendency to align with them, in competition with noise. 
While forbidden in equilibrium \cite{Mermin1966} the Vicsek model shows 
true long-range order in two dimensions \cite{Toner1995}, meaning that activity makes orientational correlations robust 
against noise over arbitrarily long distances.
%The ordered phase, called Toner-Tu phase, shows global polar order, 
%despite the presence of giant density fluctuations leading to large spatial regions which are almost void of particles. 
The phase transition from the disordered phase which occurs for strong noise to the long-range ordered Toner-Tu phase 
is now known to be discontinuous \cite{Chate2004} and features a remarkably large coexistence region \cite{Caussin2014,Solon2015}
where high-density bands of comoving polarized particles spontaneously emerge and traverse through a background of a
low-density disordered gas-like phase. 
% ordered and disordered structures coexist, similar as for a liquid-gas transition . Remarkably, the ordered structures take the form of  
% high-density bands of comoving polarized particles, which traverse through a background of a
% low-density disordered gas-like phase. 
%Similar bands occur also in other forms of polar active matter, e.g. in experiments with  or
%in chemorepulsive active colloids \cite{Liebchen2015,Liebchen2017}.
These bands behave highly randomly;
they merge when colliding with each other but also split up frequently, rendering an irregular pattern of sharply localized and strongly
polarized moving bands. 
The latter choose their direction of motion spontaneously depending on initial state and fluctuating molecular
environment (noise realization); thus, when averaging over many realizations, there is no net motion.
%reflecting the fact that isotropy is spontaneously broken by the emegence of 
%polar order from short-ranged alignment interactions. 
This randomness 
is unfortunate in view of potential key applications of active matter, e.g. for targeted drug delivery, crucially
requiring schemes to control active particles. Here, while single particle guidance with external fields is among the 
most explored problems in active matter
\cite{Palacci2013,Koumakis2013,Ma2015,Ebbens2016,Geiseler2016,Demirors2017,Colabrese2017,Geiseler2017,Muinos2018} and there 
is a moderate knowledge on 
interacting particles in external fields (complex environments) \cite{Drocco2012,Ginot2015,Kuhr2017,Reichhardt2017a,Zhu2018,Reichhardt2018b} and their control \cite{Yan2016,Peng2016,Guillamat2016,Guillamat2017,Kaiser2017},
surprisingly little is known about the controllability 
of polar active particles and band patterns they naturally form. 
%of band patterns naturally occurring in polar active matter. 

In the present work we ask for a scheme to tame band patterns, i.e. if we can force the bands in the Vicsek model to settle down into 
a pattern featuring a predictable and externally controllable direction of motion.  
To achieve this, we apply a ``traveling wave potential'' (also called travelling potential ratchet \cite{Reimann2002}) to the Vicsek model. 
We find that such an external field does in fact allow to control the late time direction of motion of particle ensembles 
in polar active matter, yet, in a remarkably nontrivial way. 
In our simulations, for appropriate parameter regimes, we see the formation of
bands that at early times pin to the 
minima of the travelling potential and comove with it. When time proceeds, 
one of the bands suddenly unpins and starts counterpropagating in the travelling potential. 
Upon subsequent collisions the band swells towards a macroband containing most particles in the system. 
This macroband emerges representatively in a large parameter window and shows a predictable motion against the 
travelling direction of the potential.
Our results show that a moving (or tilted, see Fig.~\ref{fi:setup}) substrate 
tames the collective behaviour of polar active particles and can be used 
to control the transition from microphase separation (band patterns) to 
a macrophase separated state which does show predictable transport. 
In the following, we specify these results and analyze the mechanism underlying the emergence of a counterpropagating 
macroband. 

\section{Model}
\begin{figure}[t]
\centerline{\includegraphics[width=\linewidth]{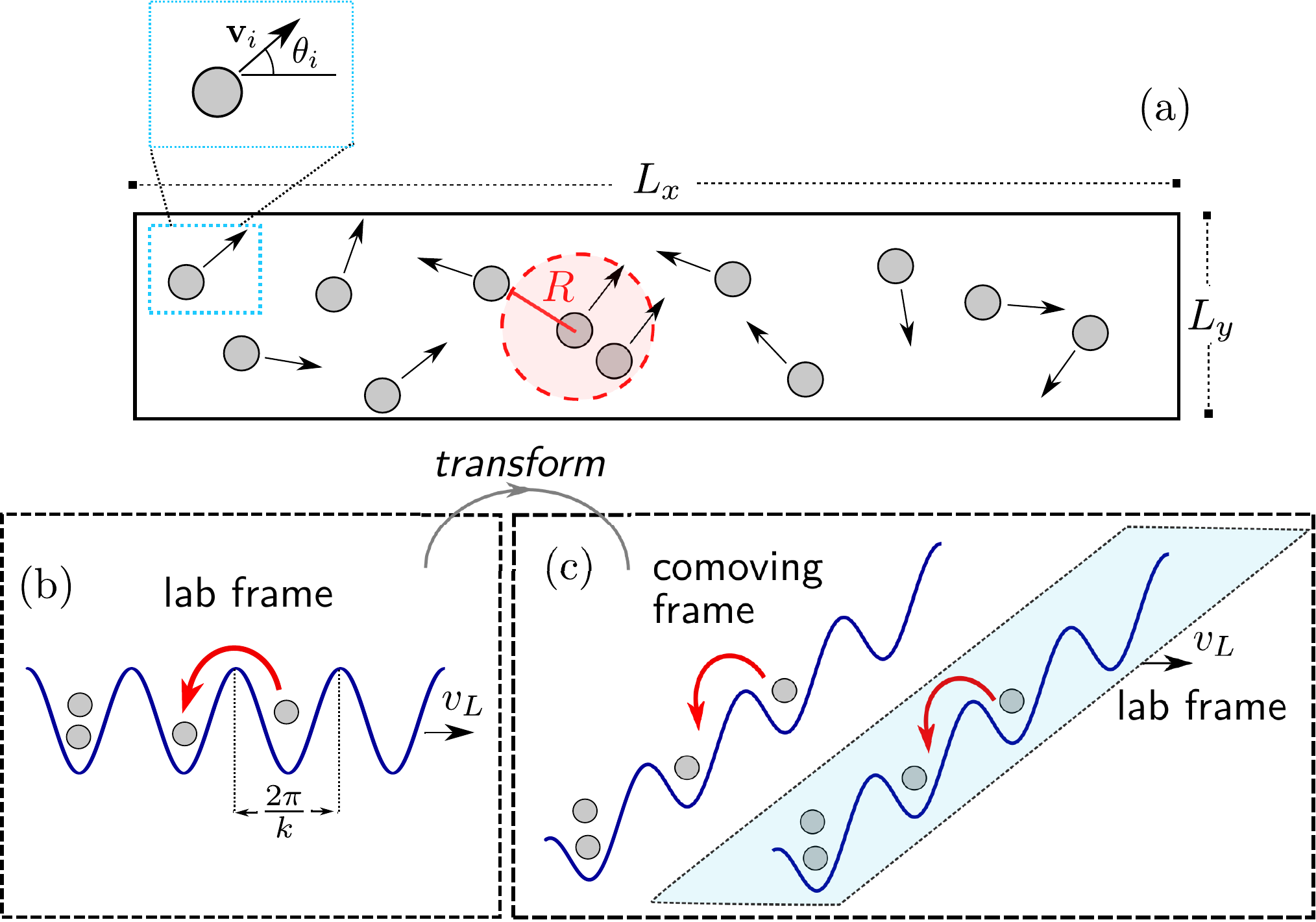}}
\caption{(a) Cartoon of the polar active particles (a) in a traveling wave potential (b), with a velocity $v_L=\omega/k$.
Here $\vec{v}_i=v_0 \vec{p}_i$ is the self-propulsion velocity of the $i$-th particle, aligning with adjacent partices (red circle). 
A Galilei transformation to the comoving frame
turns the travelling wave potential (panel (b))
into a static, tilted periodic potential (panel (c)). Thus, motion in the travelling wave potential (b) 
is equivalent to motion in a static tilted lattice in the comoving frame which displaces 
through space (relative to the laboratory frame) with a constant speed $v_L$ (panel (c)). The pinned state,
where particles comove with the travelling wave corresponds to particles resting around a minimum of the tilted lattice in (c).
We find that in the lab frame, for certain values of the parameters, polar active particles can move faster down the tilted lattice (to the left) 
than the lattice displaces through space (see Sec IV). 
The dynamical pathway to achieve this sliding state  involves a controllable transition from microphase separation (patterns)
to macrophase separation (counterpropagating macroband).}
\label{fi:setup}
\end{figure}

%Fig 1: Cartoon of the lattice and perhaps the mechanism (not needed now)?
We consider $N=5000$ active overdamped particles in a quasi-1D-potential $V(x,y,t)$, which is uniform in $y$-direction and represents a traveling wave in $x$-direction; i.e. it is 
periodically modulated and  
moves with constant speed $v_L=\omega/k$ in $x$-direction, where $\omega,k$ are the frequency and wave vector 
of the travelling potential (Fig. \ref{fi:setup} (b)).
Such a potential has previously been considered for 
active point particles \cite{Zhu2018} and disks \cite{Sandor2017} and can be realized e.g. by 
a micropatterned ferrite garnet film substrate \cite{Tierno2016}, by optical lattices traversing at speeds of a few $\mu/s$, or effectively (see Fig. ~\ref{fi:setup} (c)), 
simply by a tilted washboard potential \cite{Hanggi2009,Juniper2016,Juniper2017}.
Note here that in the comoving frame (moving with a constant velocity $v_L$) the dynamics translates into motion of a particle in a static tilted 
washboard potential (see Fig. \ref{fi:setup} (b),(c) and Sec. IV).
%the particle dynamics in the travelling wave potential is equivalent to that in a tilted washboard, despite a 
%constant displacement, see Fig.~\ref{fi:setup} (b),(c). 
\\Besides experiencing the external potential, the 
active particles also self-propel, a fact effectively described by a self-propulsion force $\gamma v_0 {\bf p}_i$ where
$\vec{p}_i=\cos \theta_i \vec{e}_x+ \sin \theta_i \vec{e}_y; \; i=1,..,N$ 
are the self-propulsion directions of the particles and $\gamma$ is the Stokes
drag coefficient. In  bulk, the particles would move with a constant speed $v_0$. 
%Hydrodynamic flow fields , phoretic 
%or steric repulsions for rod-shaped particles, can all align the particles. We here 
As in the Vicsek model, we assume that the particles align with each other.
%, which may in general occur e.g. due to hydrodynamic flow fields produced by the swimming particles, 
%phoretic fields in autophoretic colloids or due to steric interactions of rod-shaped particles. 
%When the particles are non-isotropic, they generically experience alignment interactions; here, we model such interactions 
%using the generic form as in the Vicsek model 
We define the dynamics of the particles by the following equations 
of motion \cite{footnote2}:
\begin{subequations}
\begin{eqnarray}
\dot {\bf r}_i &=& v_0 {\bf p}_i + {\bf F}_i/\gamma \label{emot1a} \\
\dot{\theta}_i &=& \frac{g}{\pi R^2}\sum_{j \in \delta_i}^{N}\sin\left(\theta_j-\theta_i\right)+\sqrt{2D_r}\eta_i(t) \label{emot1b} 
%\dot{x}_i &=& v_0\cos \theta_i+U_0\cos\left(2 \pi \left(kx_i-\omega t\right)\right)   \label{emot1b} \\
% \dot{y}_i &=& v_0\sin \theta_i, \label{emot1c}
\end{eqnarray}
\end{subequations}
Here, $g$ controls the strength of alignment of a particle with its neighbors within a range $R$ and the sum is performed over all these
neighbors (see Fig. \ref{fi:setup} (a)). Alignment competes with rotational Brownian diffusion, occurring with a rate $D_r$; 
$\eta_i$ represents Gaussian white noise of zero mean and unit variance. 
The force due to the substrate reads
\begin{equation}
{\bf F}_i:=-\nabla U=\gamma u_0\cos\left(2 \pi \left(k x_i-\omega t\right)\right){\bf e}_x \label{lat1}
\end{equation}
% \1
% {\bf f}_i:=-\nabla U/\gamma=u_0\cos\left(2 \pi \left(k x_i-\omega t\right)\right){\bf e}_x
% \2
where $u_0$ is the strength of the external force. %$u_0=U_0/\gamma$ is the height of the potential reduced by the friction coefficient of the particles. 
Here, in all of our results we express lengths and times in units of $\mu m$ and $s$ respectively, 
i.e. we introduce parameters $D_r'=D_r \cdot s,~ g'=g\cdot s/\mu m^2$ etc. and omit primes for simplicity,
allowing thus for a straightforward comparison with potential experiments.
However, for readers interested in the actual dimensionless control parameters, see \cite{footnote1}.

We now study the dynamics of the described model using Brownian dynamics simulations and 
an elongated simulation box of size $L_x \times L_y = 500 \times 5$, fixing the density to $\rho=2$, 
as well as random but uniformly distributed initial particle positions and orientations. 
Using more quadratic boxes leads to qualitatively similar phenomena.
%we express all of our results in units of $1\mu m$ (for length) 
%and $1s$ (for time) 

\section{Counterpropagating Macroband} %Phenomenology

\begin{figure*}[t]
\begin{center}
\includegraphics[width=0.76\linewidth]{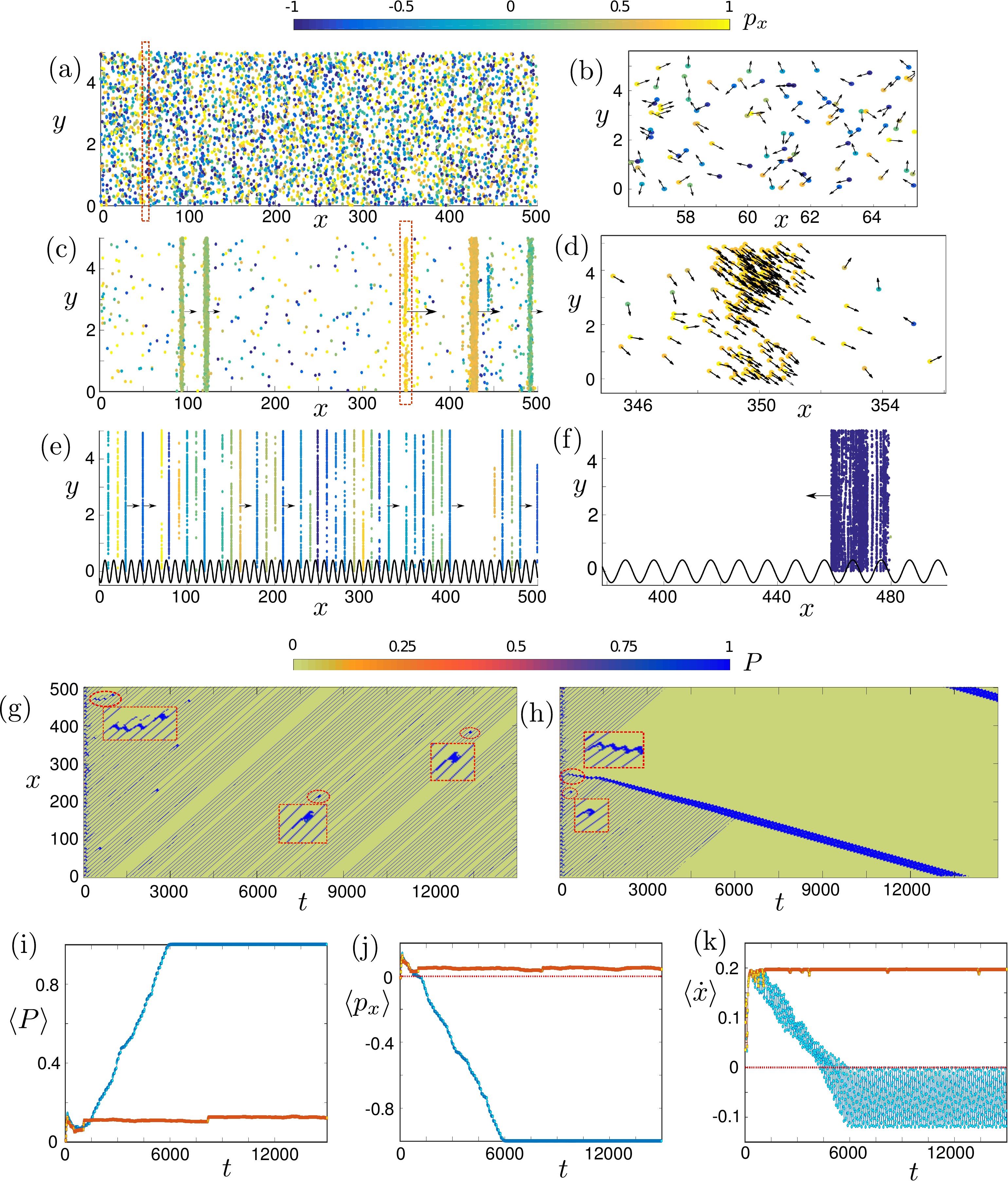}
\end{center}
\caption{(a),(b),(c),(d) Snapshots of (a) the disordered uniform phase and (c) the bands formed in the ordered phase in the absence of the lattice  from
sample simulations at $D_r=0.151$ and
$D_r=0.003$ respectively. The corresponding zoomed-in figures (b),(d) demonstrate the direction of the particles' motion. 
(e),(f) Snapshots of the two different ordered phases: (e) pinned phase and (f) sliding phase in the presence of a lattice from sample simulations at $D_r=0.0003$. 
Note that the shown 'sine wave' illustrates the lattice focusing on its wavelength.
In all the above cases (a)-(f) the colors denote the polarization of each particle along the $x$ direction
$p_{x,i}=\cos \theta_i$.(g),(h) The polarization $P_n$ of the formed bands (Eq. (\ref{polarization})), depicted by color, as a function of the time and the $x$ coordinate
for sample simulations of  (g) the pinned phase and (h) the sliding phase at $D_r=0.0003$. The insets provide zooms of some special events during dynamics featuring  band collisions.
(i)-(k) Time evolution of (i) the average polarization $\avg{P}$, (j) the average cosine of the particles $\avg{p_x}=\avg{\cos \theta_i}$
and (k) the average particle velocity $\avg{\dot{x}}$ for the pinned (dark orange line) and the sliding phase (blue line) of (g),(h). 
In the above cases the parameters of our setup read $g=0.07$, $D_r^c=0.15$, $N=5000$, $L_x=500$, $L_y=5$, $v_0=0.2$ and in most cases $u_0=0.3,\omega=0.02,k=0.1$ ($v_L=0.2$). }
\label{fi:cu_rev}
\end{figure*}

In the absence of a lattice, our simulations reveal the usual phenomenology of the Vicsek model \cite{Chate2004,Caussin2014,Solon2015}: 
For a given alignment strength ($g=0.07$) and comparatively strong noise 
$D_r>D_r^c\approx 0.15$ (or high temperatures), we find a disordered uniform phase 
(Fig. \ref{fi:cu_rev} (a),(b)), whereas noise values $D_r<D_r^c$ lead to a 
polarized phase (Fig. \ref{fi:cu_rev} (c),(d)). 
In the latter phase, particles self-organize into polarized bands of high density which move with a speed $\sim v_0$ and 
coexist with gas-like unpolarized regions in between the bands. 
The bands occur at seemingly irregular distances to each other. As time proceeds, they 
occasionally split up (for $D_r\neq 0$) and typically merge when they collide with each other; 
overall, the number and size of the bands changes dynamically. 
%regime where particles are not polarized in between. 
%When colliding with each other, the bands typically merge. They sometimes also slip up spontaneously (unless 
%$D_r=0$); so that 
%even at late times we have a state where the number of bands continuously increases and decreases. 

In the presence of the traveling wave (moving lattice) 
and at weak noise ($D_r=3\times 10^{-4}<D_r^c$) the behaviour of the bands may change dramatically. 
While very steep lattices of course pin the particles permanently to the lattice minima, leading to a state where all particles comove with the lattice, 
shallow lattices have little impact on the behaviour of the system and its tendency to form bands. 
In this latter regime, the lattice exerts a 
periodic force which essentially averages out before the particles move much. 
Thus, we here focus on moderate lattice 
depth ($u_0=0.3$) and lattice speeds comparable to that of the particles ($v_0=v_L=0.2$), so that the particles can occasionally overcome the potential maxima.
In this regime, for sufficiently weak noise (here $D_r<D_r^c=3\times 10^{-4}$), particles form quickly bands, 
most of which are pinned to the lattice and thus co-move with it (Fig. \ref{fi:cu_rev} (e),(g),(h)). Note that the polarization of 
such bands 
\begin{equation}
P_n = \sqrt{\left( \sum_{i \in \textrm{band}~n} \cos \theta_i \right)^2+\left(\sum_{i \in \textrm{band}~n} \sin \theta_i \right)^2} \label{polarization}
\end{equation}
is almost unity even for small times and maintains this very high value during the time evolution  (Fig. \ref{fi:cu_rev} (g),(h)).
Occasionally, we observe that a band, assisted by the existing noise, changes direction and counterpropagates; it then soon collides with another
band (see Fig. \ref{fi:cu_rev} (g),(h) insets for such collision events).
Here, the two bands merge and form one larger band which in some cases becomes pinned and in other cases slides, still against the direction of motion of the lattice.  
In the latter case, the band soon encounters further bands and can in each case, either stop moving (get pinned to the lattice) or continue sliding. 
One might expect that this seemingly random result of the collision processes should ultimately lead back to a pinned state. 
Strikingly, however, in many simulations we observe cases where
a band counterpropagates through the entire lattice and systematically consumes all other bands. 
The result is one macroband which contains most of the $N$ particles and counterpropagates against the direction of lattice motion (Fig. \ref{fi:cu_rev} (f),(h)). 
Since the particles counterpropagate, even when viewed from the laboratory frame, with respect to the forces acting on a pinned particle in a minimum of the lattice, they feature an 
absolute \textit{negative mobility}.
Thus, we observe a spontaneous reversal from a comoving state where most particles have followed the lattice to a counterpropagating state. 
% Depending on the particle and noise configuration, this process can go on and on with the resulting band colliding successively with all other existing bands 
% and absorbing them to a large macroband that counterpropagates in the lattice (Fig. \ref{fi:cu_rev} (f),(h)). 
% Thus, within the aforementioned parameter regime there is the possibility of a spontaneous current reversal with a final state 
% in which almost all particles form a single macroband moving opposite to the lattice (Fig. \ref{fi:cu_rev} (f)). 

The striking difference between a finally pinned (Fig. \ref{fi:cu_rev} (e),(g)) and a finally sliding state (Fig. \ref{fi:cu_rev} (f),(h))
featuring a current reversal is further illustrated in Fig. \ref{fi:cu_rev} (i),(j),(k). Here we observe that the mean polarization (averaged over all particles)
$\avg{P}=\sqrt{\avg{\cos \theta}^2+\avg{\sin \theta}^2}$ increases from the pinned state at short times to a value of almost one 
for the sliding macroband (Fig. \ref{fi:cu_rev} (i)).
It turns out (Fig. \ref{fi:cu_rev} (j))
that $\avg{p_x}=\avg{\cos \theta} \approx -1$, meaning that the particles collective self-propel against the direction  
of the lattice motion (still in the laboratory frame), i.e. along $-\vec{e}_x$. The average velocity of the particles is $\avg{\dot{x}}\approx v_L=0.2$ (see also Eq. (\ref{emot1a}))
for the pinned state (Fig. \ref{fi:cu_rev} (k)) and 
acquires a negative value oscillating in time for the case of the sliding macroband.

% When slightly enhancing the noise strength $D_r\gtrsim D_r^c$, we observe that the 
% particle current initially follows the one for weaker noise, but then, strikingly, from $t\sim 1000$ onwards 
% slows down and finally even 
% reverts its direction. That is, noise controls the late-time transport direction and can induce a reversal of the 
% transport direction.
% On the level of the inidividual bands, we find that after a collision of a co- and a counterpropagating band, 
% the system does not always
% return to the globally pinned state, but two bands may continue sliding against the direction of the lattice
% motion (Fig 2j).
% At some point, one of the bands counterpropagates through the entire lattice and systematically consumes 
% all other bands (Fig. 2i). At late times, all particles form a single macroband moving opposite to the lattice.

\section{Pinned and sliding solutions} %Analysis -- the Mechanism
We can get some first insight into the mechanism underlying the surprising counterpropagation of the bands by 
examining the single-particle dynamics in the zero noise limit. When projected to the $x$-axis Eq. (\ref{emot1a}) reduces to 
\begin{equation}
\dot{\tilde{x}}={\tilde{v}_x}+\tilde{u}_0 \cos \tilde{x},\label{pendulu1}
\end{equation}
where the Galilean transformation to the comoving  frame $\tilde{x}=2 \pi \left(kx-\omega t\right)$ is used, with ~$\tilde{v}_x=2 \pi k \left(v_0 p_x-v_L\right)$,~$-1\leq p_x=\cos \theta \leq 1$ and
$\tilde {u}_0=2 \pi k u_0$. 
This equation is well known 
as the overdamped limit of the equations of motion of e.g. the forced nonlinear pendulum \cite{Coullet2005},
the driven Frenkel-Kontorova (FK) model \cite{FKbook1} and the resistively shunted junction (RSJ) model of Josephson junctions \cite{Barone1982}. It is known
to attain two different kinds of solutions depending on the value of $\frac{\tilde{v}_x}{\tilde{u}_0}$.  For $\abs{\frac{\tilde{v}_x}{\tilde{u}_0}}  \leq 1$ the system 
is in the so-called \textit{pinned phase} where the particle cannot overcome the 
potential barrier $\tilde{u}_0$ and remains therefore trapped within one of its wells  
($\avg{\tilde{x}}_t^{\infty} \rightarrow 0)$, yielding an asymptotic time averaged velocity $\avg{\dot{x}}_t^{\infty}=v_L=\omega/k$. 
In the opposite case $\abs{\frac{\tilde{v}_x}{\tilde{u}_0}} > 1$, the particle is fast enough 
to overcome the potential barrier 
separating the wells (or in the example of the pendulum to lead to a rotation) 
and thus the system exhibits a \textit{sliding phase} where the particle
permanently moves  (slides or rotates) in one and the same direction with an oscillating velocity $\dot{x}$ \cite{Cheng2015} of period 
%\begin{equation}
$T=\frac{2\pi}{\sqrt{{\tilde{v}_x}^2-\tilde{u}_0^2}}$ \cite{Cheng2015} %\label{period_sliding}
%\end{equation}
and an asymptotic time averaged velocity
\begin{equation}
\avg{\dot{x}}_t^{\infty}=\textrm{sgn} \left(v_0 p_x-v_L\right)\sqrt{\left(v_0 p_x-v_L\right)^2-u_0^2}+v_L. \label{vel_sliding} 
\end{equation}
%These considerations elucidate  the long time behaviour of the average velocity of the  many body system in the two different states (Fig. \ref{fi:cu_rev} (k)).

For the $N$-particle system (Eqs.(\ref{emot1a}),(\ref{emot1b})) the particles' self-propulsion directions $\vec{p}_i$
change due to alignment interactions (Eq. (\ref{emot1b})) and
noise (Eq.(\ref{emot1a})). Hence, the projection of the particle speed onto the $x$-axis changes in time, so that 
the sliding condition 
$\abs{\frac{\tilde{v}_x}{\tilde{u}_0}} > 1$ subsequently may and may not be fulfilled. 
In terms of $\vec{p}_i$, the sliding condition reads
\begin{equation}
-1< {p_x}_i = \cos \theta_i<\frac{v_L-u_0}{v_0} ~~\textrm{or}~~  1>{p_x}_i >\frac{v_L+u_0}{v_0}. \label{cond1}
\end{equation}
In the present parameter regime ($v_0=v_L=0.2,~u_0=0.3$) 
sliding occurs for ${p_x}_i <-\frac{1}{2}$,
or $\theta_i \in \left[\frac{2\pi}{3},\frac{4\pi}{3}\right].$ Thus
roughly $1/3$ of particles will initially be in the sliding phase. Importantly, all of these particles which can in principle slide, move \textit{against} the direction of lattice motion (negative ${p_x}_i$), i.e. 
sliding can only occur against the direction of lattice motion, as observed 
%if they encounter a lattice barrier before their direction changes. 
%will have the possibility to slide, in accordance to what observed
in Figs. \ref{fi:cu_rev} (f),(h),(j),(k). 
The main effect of rotational diffusion (noise in the particle orientations) 
consists in the smoothening of the pinned-to-sliding transition at $\abs{\frac{\tilde{v}_x}{\tilde{u}_0}}= 1$.%, as can be seen based on the Smoluchowski
%Not sure which noise you meant here..
%Note that in the presence of white noise Eq. (\ref{pendulu1}) obtains the form of the well known Smoluchowski equation \cite{Cheng2015}

\begin{figure*}[t]
\centerline{\includegraphics[width=0.78\linewidth]{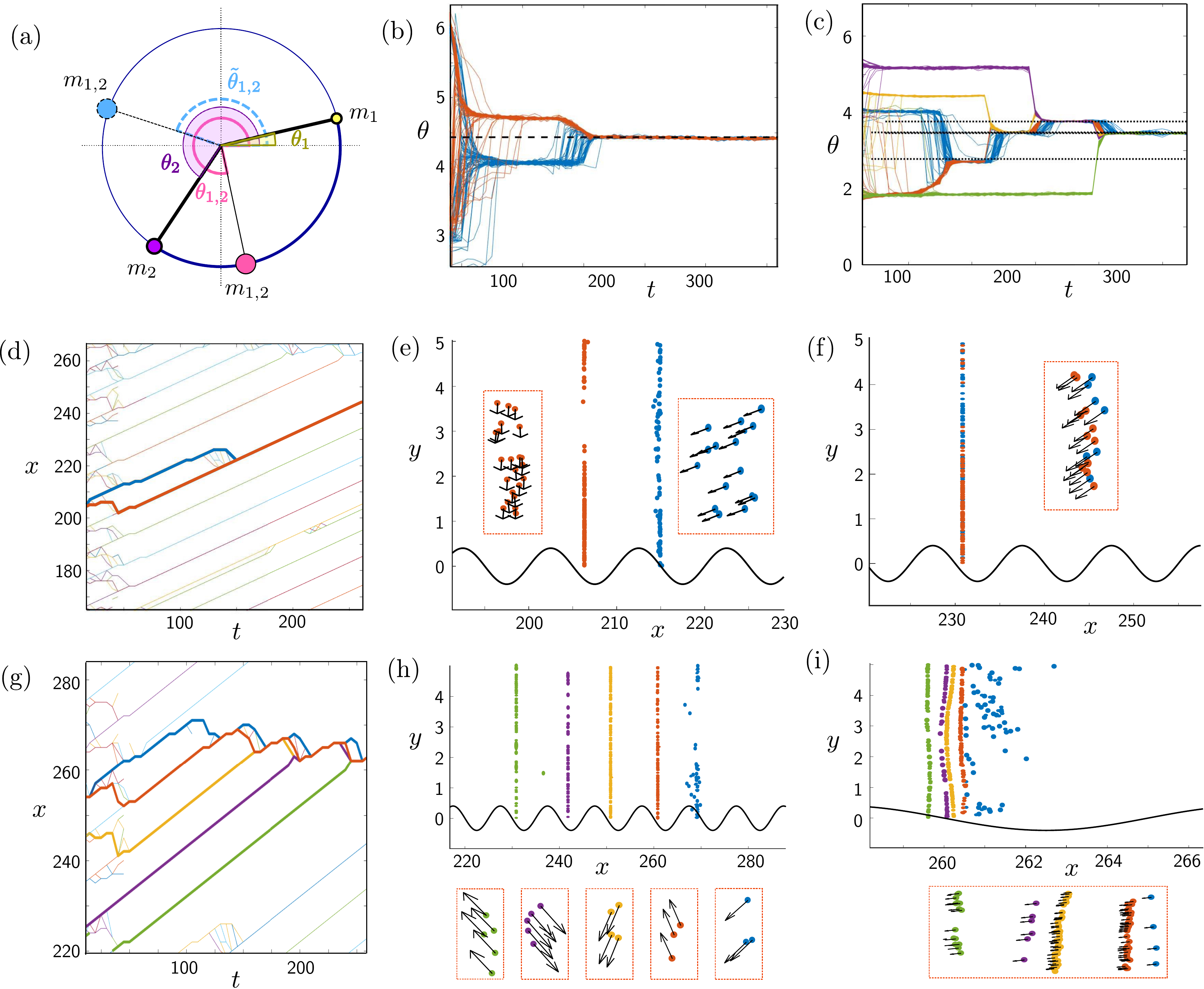}}
\caption{
(a) Schematic illustration of the mechanism of merging for two bands with  angles $\theta_1$ and $\theta_2$ and masses
$m_1$, $m_2$ leading to the formation of a large band with angle $\theta_{1,2}$ (stable) or $\tilde{\theta}_{1,2}$ (unstable).
(b) Time evolution of the angles $\theta$ of two colliding bands which after their merging become pinned in the lattice (see also subfigures (d)-(f)). 
(c) Time evolution of the angles $\theta$ of five bands 
colliding successively with each other, leading to the formation of a large sliding band (see also subfigures (g)-(i)).
(d) Time evolution of the $x$ coordinate of two bands which after their merging become pinned in the lattice. Figures (e) and (f) provide corresponding snapshots before and after
the collision, respectively. 
(g) Time evolution of the $x$ coordinate of five bands which successively collide and merge, initiating the formation of the large sliding band.  Figures (h) and (i) provide corresponding snapshots before the first and after
the final collision, respectively. In the above cases the parameters of our setup read $g=0.07$, $D_r^c=0.15$, $N=5000$, $L_x=500$,$L_y=5$, $v_0=0.2$,$D_r=3 \times 10^{-4}$, $u_0=0.3,\omega=0.02,k=0.1$ ($v_L=0.2$).}
\label{fi:col1}
\end{figure*}

\section{Collisions of Vicsek Bands}
% As mentioned above, the alignment interactions of the Vicsek model (Eq. (\ref{emot1b})) lead for a quasi 1D setup ($L_x \gg  L_y$)
% and low noise ($D_r<D_r^c$) to the formation of bands along the $y$ direction (Fig. \ref{fi:cu_rev} (c),(d)). 
% This effect is enhanced in the presence of 
% the lattice along $x$ with the different bands occupying for short times individual wells. 
We now exploit these considerations regarding pinned and sliding states for single particles to understand the dynamics of the polarized bands in the lattice. 
%now discuss collision events of bands 
In our simulations, shortly after their formation, the magnitude of the polarization of the individual bands quickly approaches a value close to one 
(Fig. \ref{fi:cu_rev} (g),(h)), i.e. most bands are moving with almost constant individual velocities relative to the lattice.
Thus, in the absence of collisions, the bands essentially behave like single particles and are either pinned or slide through the lattice. 
To study band collisions, it is useful to assign effective ``masses'' $m_n$ to the bands representing the number of particles contained in the band.
%\textcolor{red}{(up to an aribitrary multiplicative constant) ? Why a ultipicative constant??}.  
%The presence of interaction and noise alters this picture, affecting crucially the dynamics of the system 
%through  the emergence of band collisions (Fig. \ref{fi:cu_rev} (g),(h) (inset)). After 
 When two bands with polarization angles $\theta_1, \theta_2$ and masses $m_1, m_2$ 
collide, they usually merge into a larger band of total 
mass $m_{1,2}=m_1+m_2$ (Fig. \ref{fi:col1}(a)) and average their polarizations.
(Formally, there are two fixpoints of the orientational dynamics when two bands merge: one reads 
$\theta_{1,2}= \Theta_0$ with $\Theta_0=\frac{m_1 \theta_1+m_2\theta_2}{m_1+m_2}$, the other one 
$\theta_{1,2}=\Theta_0-2\pi \frac{m_1}{m_1+m_2}$; here the one lying within the smaller arc 
between $\theta_1$ and $\theta_2$ is stable and thus observed, the other one is unstable.)

In our simulations, the polarization direction of an isolated band can freely rotate (Goldstone mode); noise therefore creates
a random dynamics of the band polarization direction. Once the polarization angle of an initially pinned band reaches a value 
$\theta_1 \in (\frac{2\pi}{3},\frac{4\pi}{3}$) the band will move over a lattice barrier in the direction opposite to the lattice motion (Figs. \ref{fi:col1}(d),(e),(g),(h)). 
%This motion occurs when the band reaches a polarization direction which is only marginally sliding, i.e. $\theta_1$ is close to $2\pi/3$ or $4\pi/3$.
Since the motion of the band towards an adjacent lattice site occurs on timescales which are short compared to the time noise needs to significantly change $\theta_1$, 
the band will typically feature an angle close to $2\pi/3$ or $4\pi/3$ when it encounters another pinned band. 
Depending on its relative orientation to the band it encounters, after the collision, its angle may either be out of the sliding interval (Fig. \ref{fi:col1}(b)), 
\emph{or, similarly likely, may be deeper in the sliding interval} (Fig. \ref{fi:col1}(c)). In the latter case, the band continues counterpropagating through the lattice. 
Statistically, further collisions with other bands can be essentially viewed as a random walk of the band's polarization direction. Here, however, the 
effective mass of the band increases within each collision, corresponding to a decrease of the stepsize after each step. 
% {\color{red} Comment: The change in phase of the macroband can be thought of as a random walker with decreasing stepsize (stepsize $1/n$ at step $n$). The lifetime (distribution) of the macroband 
% in a very large system can be estimated by initiating such a random walker in the centre of the sliding interval and calculating the mean first passage time of one of the boundaries of the 
% sliding interval. (This random walker will not feature a fixpoint of course, so the lifetime should be finite - but probably very large). 
% We are still thinking about this point}
Hence, when the polarization of 
a band after a first few collisions is deeply in the sliding regime, i.e. $\theta \approx \pi$, the sliding of the band is highly robust against further collisions. 
This is why we have observed the emergence of a counterpropagating macroband consuming all other bands
(Fig. \ref{fi:cu_rev} (f),(h)).

To understand the broad width of the counterpropagating macroband, it is instructive to resolve the collisions slightly further. 
When sliding bands collide, the positions of the contained particles do not fully mix up; rather, the resulting band features a 
substructure of microbands stacked one behind the other
(Fig. \ref{fi:col1} (i)). This fact is responsible for the large width of the observed macroband (Fig. \ref{fi:cu_rev} (f)) in the case of  finally sliding states.
This is because successive collisions typically result in a macroband with an average polarization close to $\pi$, which is the centre of the sliding interval $\left[\frac{2\pi}{3},\frac{4\pi}{3}\right]$.
Thus after the collisions the involved particles move 
in the negative $x$ direction with $\vec{p}_i \approx -\vec{e}_x$ (see Fig. \ref{fi:col1} (i)), a fact that prohibits them from mixing along the $y$ direction.
Conversely, band collisions leading to pinning do not induce a pronounced substructure. 
Here, the involved particles move significantly in the $y$-direction and therefore 
tend to mix in the course of the dynamics (Fig. \ref{fi:col1} (f)).

\section{Effect of the particle speed}
\begin{figure}[t]
\centerline{\includegraphics[width=\linewidth]{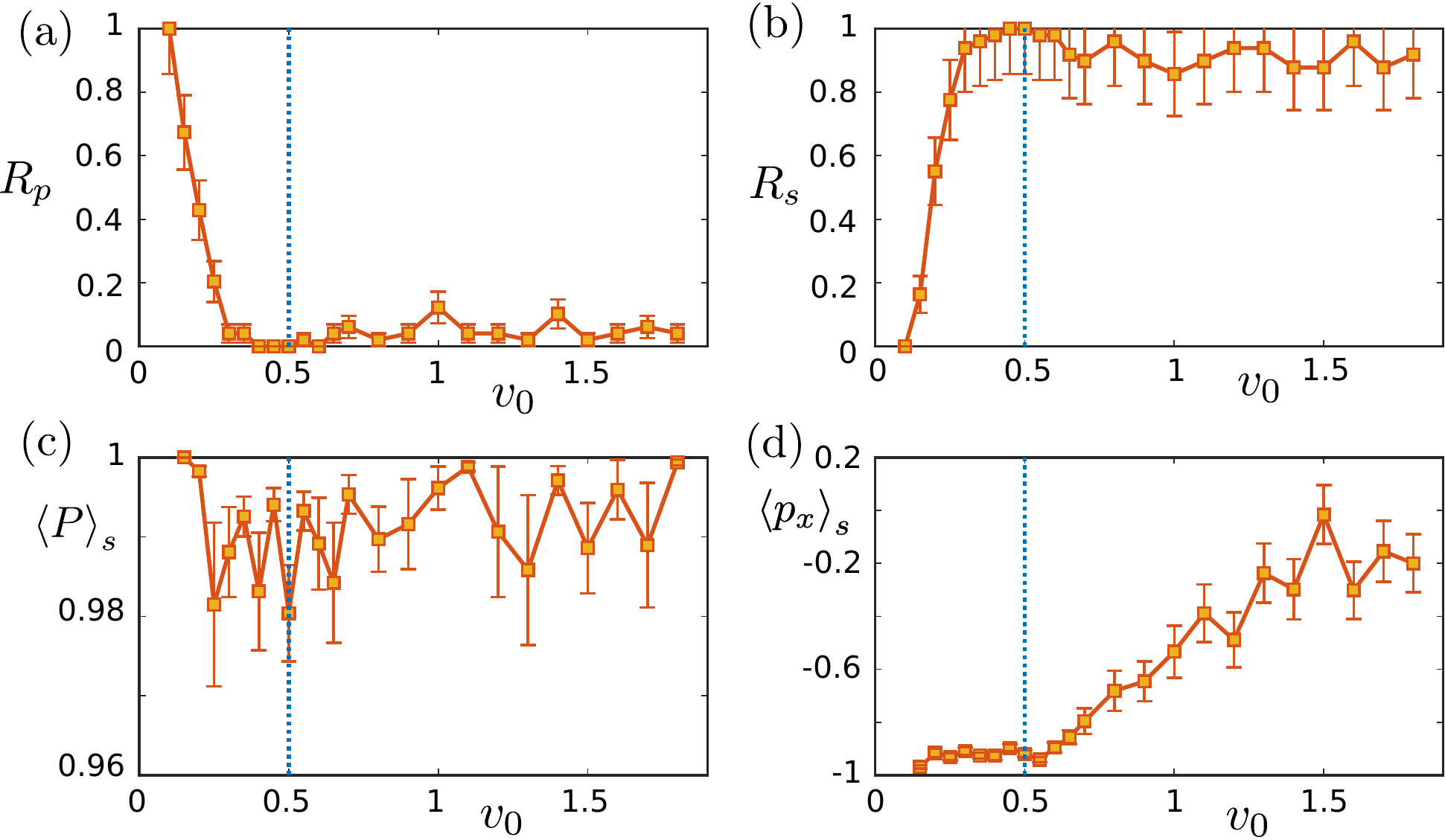}}
\caption{(a) Ratio of the finally pinned states $R_p$(lower bound) as a function of the particle velocity $v_0$.
(b) Ratio of the finally sliding states $R_s$ (lower bound) as a function of the particle velocity $v_0$. 
 (c) Average polarization of the bands in the finally sliding states $\avg{P}_s$ as a function of $v_0$.
(d) Average cosine of the bands $\avg{p_x}_s$ in the finally sliding states  as a function of $v_0$. The inset shows the respective standard deviation of ${p_x}…_s$
for the different realizations.  Here we have used $g=0.07$, $D_r^c=0.15$, $N=5000$, $L_x=500$,$L_y=5$, $D_r=3 \times 10^{-5}$, $u_0=0.3,\omega=0.02,k=0.1$ ($v_L=0.2$).
The dotted vertical line marks the value of the velocity $v_0=0.5$ beyond which sliding in the positive direction is possible.}
\label{fi:v0fig}
\end{figure}

Having explored the mechanism leading to the dynamical reversal of the direction of motion of the particles in the moving lattice, 
we now ask how representative this scenario is. 
 Here, we stay in the low noise regime ($D_r=3 \times 10^{-5}$) and with our previous values for the lattice velocity $v_L=0.2$ and height $u_0=0.3$ but vary the self-propulsion velocity of the 
 particles. For $v_0 \leq 0.1$, we have $u_0-v_L>v_0$ and hence sliding is not possible, whereas for 
$v_0>0.5$, where $v_0>u_0+v_L$, sliding can be achieved also in the positive direction (see Eq. (\ref{cond1})). In the complete interval 
$0.1<v_0 \leq 0.5$ sliding is possible only against
 the lattice motion (negative direction) as discussed above. Within this interval, larger values of $v_0$ yield a larger interval of polarization angles 
 leading to sliding (Figs. \ref{fi:v0fig} (a),(b)). 
 To specify this, we simulate 50 particle ensembles for each value of $v_0$ and count the number (ratio) of ensembles $R_s$ which 
 have reached a sliding and counterpropagating macroband and the corresponding ratio of ensembles $R_p$ which have settled in an overall pinned configuration. 
Fig. \ref{fi:v0fig} (b) shows that $R_s$ increases monotonically in $v_0$, crossing from a regime where most of the bands are pinned, even at late times ($v_0\approx 0.1$)
to a regime, where $R_s \approx 1$ ($v_0\approx 0.5$). Thus, faster self-propulsion favors the emergence of a current reversal. 
(Note that the values of $R_p,R_s $ shown in Fig. \ref{fi:v0fig}(a),(b) should be viewed as lower bounds for the ratio of pinned and sliding states, as not all initial ensembles may have reached one or the 
other state at the end of our simulations.)
 
Complementary information about the finally sliding states, featuring a current reversal, is provided by Figs. \ref{fi:v0fig} (c),(d).
The final average polarization of these states $\avg{P}_s$ is for all $v_0$ very close to 1 (Fig. \ref{fi:v0fig} (c)), 
 owing to the low noise which results in particles clustering
 to a macroband with a certain alignment. In contrast, the direction of this alignment, quantified by $\avg{p_x}_s=\avg{\cos \theta}_s$, is affected strongly by $v_0$
 (Fig. \ref{fi:v0fig} (d)). For $0.1<v_0 \leq 0.5$ we have  $\avg{p_x}_s \approx -0.95$, indicating that within this interval the particles' velocities
 are all approximately aligned towards $-\vec{e}_x$ (as in Fig. \ref{fi:col1} (i)) and thus the particles counterpropagate at roughly their maximum velocity. This picture changes when 
 $v_0>0.5$ where a sliding also in the forward direction becomes possible. Different realizations result in finally sliding states 
 with a different alignment $\vec{p}_i$ and thus their ensemble average $\avg{p_x}_s \rightarrow 0$ as $v_0$  increases, recovering the isotropy in the direction
 of sliding bands of the Vicsek model in the absence of a lattice.

\section{Effect of the noise amplitude}
\begin{figure}[t]
\centerline{\includegraphics[width=\linewidth]{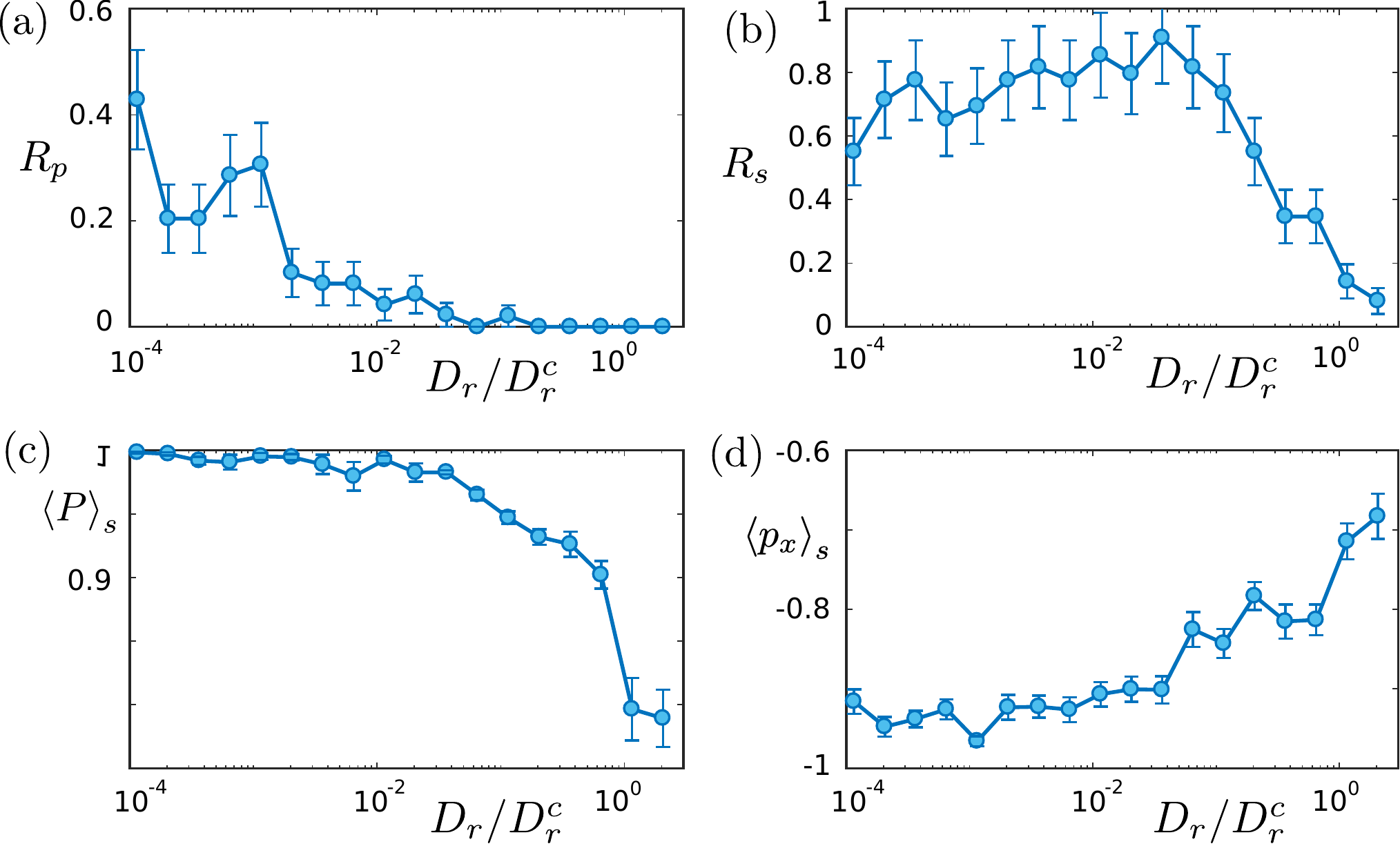}}
\caption{(a) Ratio of the finally pinned states $R_p$ (lower bound) as a function of the noise amplitude $D_r$ (in units of $D_r^c$). 
(b) Ratio of the finally sliding states $R_s$ (lower bound) as a function of the noise amplitude $D_r$ (in units of $D_r^c$). 
(c) Average polarization of the bands in the finally sliding states $\avg{P}_s$ as a function of $D_r$.
(d) Average cosine of the bands $\avg{p_x}_s$ in the finally sliding states  as a function of $D_r$. 
Here we have used $g=0.07$, $D_r^c=0.15$, $N=5000$, $L_x=500$,$L_y=5$, $v_0=0.2$, $u_0=0.3,\omega=0.02,k=0.1$ ($v_L=0.2$).}
\label{fi:Drfig1}
\end{figure}

\begin{figure}[t]
\centerline{\includegraphics[width=\linewidth]{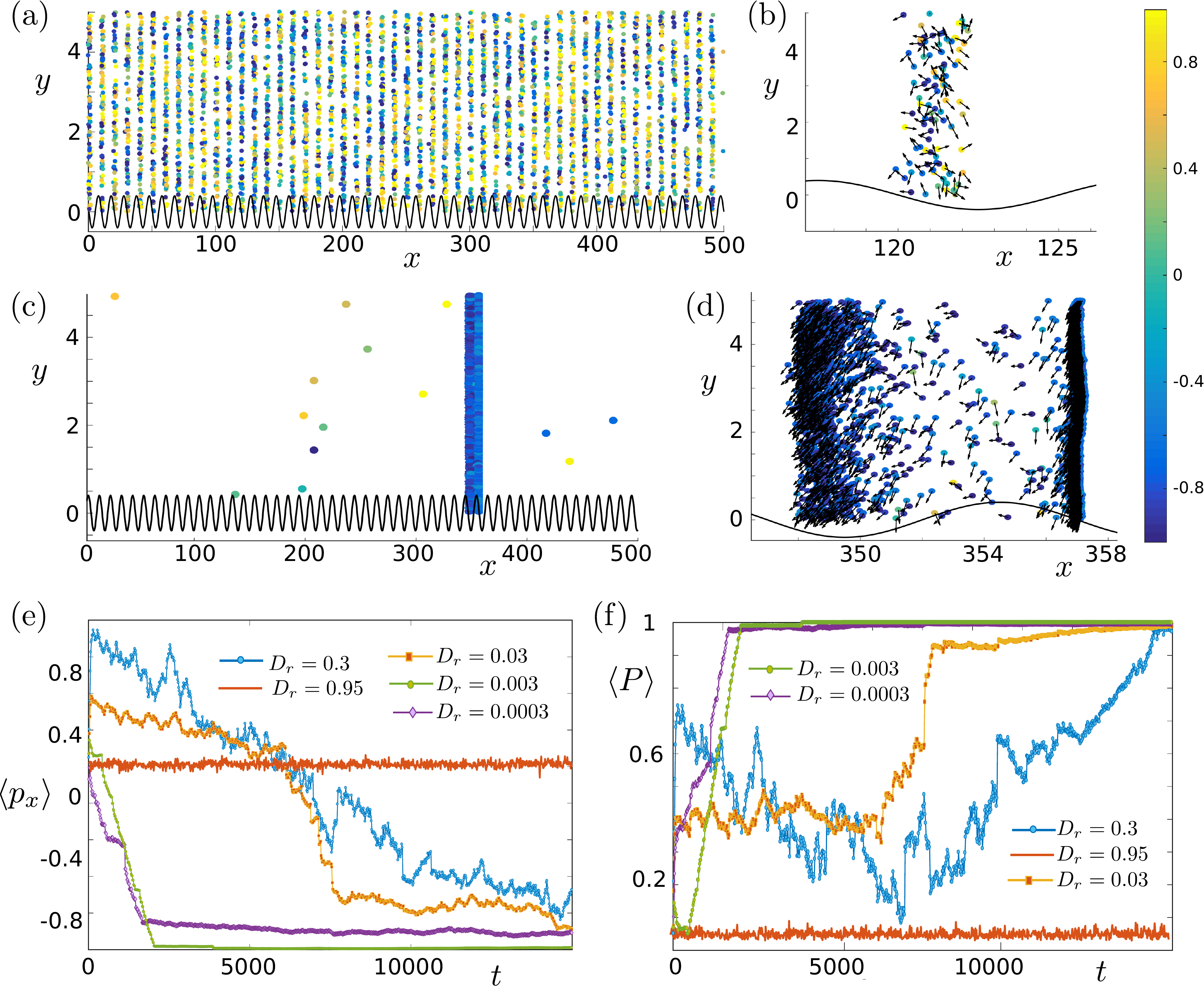}}
\caption{
(a),(b) Snapshot of the pinned phase in the presence of a lattice from a sample simulation at high noise 
$D_r=0.95$ and a zoomed figure showing the orientation of the particles.
(c),(d) Snapshot of the sliding  phase in the presence of a lattice from a sample simulation at high noise 
$D_r=0.3$ and a zoomed figure showing the orientation of the particles.
(e),(f) Time evolution of (a) the average cosine $\avg{p_x}$ and (b) the average polarization $\avg{P}$ for sample simulations leading to a
 finally sliding state for five different values of $D_r$. 
 Note the difference of Figs (c),(d) from Fig. \ref{fi:cu_rev} (f).}
\label{fi:hiDr1}
\end{figure}

Rotational diffusion, whose strength is controlled by $D_r$, is crucial to initiate the emergence of sliding states; on the other hand, there is an upper critical noise strength $D_r^c$ above which the 
Vicsek model does not show polar order, but is in the isotropic phase. 
We now systematically explore how the 
transition to the counterpropagating macroband is affected when changing $D_r$.
For $u_0=0.3,v_0=v_L=0.2$ 
where sliding is possible only in the negative $x$ direction, the ratio of states which are pinned at the end of our simulations $R_p$ decreases as 
$D_r$ increases (Fig. \ref{fi:Drfig1} (a)) and finally approaches zero for $D_r \gtrsim 0.1 D_r^c$. 
The reason for this behaviour is probably that larger noise turns the orientation of the polarization of initially pinned bands faster and therefore initiates sliding 
earlier (and more often). 
The respective ratio of finally sliding states $R_s$ (Fig. \ref{fi:Drfig1} (b)) increases only slightly as $D_r$ increases from zero for
$D_r \lesssim 0.05 D_r^c$ and afterwards decreases tending towards zero. Physically, when $D_r$ is too large, the polarization of a band may significantly change between each subsequent collisions with other bands and may 
therefore leave the sliding regime before encountering another collision. Thus, for 
too strong noise, the emergence of a transition to a counterpropagating macroband is rather unlikely. The generic behaviour of the system
for such high noise values is that of a mixture of individual particles, both in the pinned and in the sliding phase, whose polarization orientation changes fast in time,
providing the picture of an overall disordered phase modulated by the existing potential wells (Figs. \ref{fi:hiDr1} (a),(b)).
There is however still a possibility of obtaining a finally sliding state even in the high noise regime (Fig. \ref{fi:Drfig1} (b)). 
Such states are significantly less polarized than the ones in the low noise regime (Figs. \ref{fi:Drfig1} (c), \ref{fi:hiDr1} (c)) featuring also a larger variety in the direction 
of alignment, quantified by $\avg{p_x}_s$ (Figs. \ref{fi:Drfig1} (d), \ref{fi:hiDr1} (d)). Furthermore, for such cases of high noise (e.g. $D_r=0.3$, $D_r=0.95$)
the time evolution of both  $\avg{p_x}$ and $\avg{P}$ is much slower (Figs. \ref{fi:hiDr1} (e),(f))
than the ones observed for lower noise values (e.g. $D_r=0.0003$, $D_r=0.003$), indicating the diffusive character of the dynamics  expected for highly noisy systems.

% \textbf{Briefly discuss the following steps}
% \\1. Discuss mapping to pendulum equation - pinned and sliding solutions for individual particles in absence of 
% noise; briefly mention: with noise (and alignment interactions); dynamics of each particles 
% passes through stages of pinned and sliding dynamics
% \\2.1 Discuss two particle case and bistability. Conceive a cartoon visualizing them
% \\2.2 Explain why the two fixpoints make a difference in the presence of the lattice
% \\3. If particles in a band are aligned, the above single particle discussion applies to 
% dynamics of whole bands and the two-particle case describes band collisions. 
% \\3.1. Averaging collisions (illustrate with some figure from the report)
% \\3.2. Collisions leading to sliding (illustrate with some fig from report) - perhaps use Figs. 12a-d from report here. 
% \\4. Find explanation why averaging collisions are more relevant 

%%%%%%%%%%%%%%%%%%%%%%%%%%%%%%%%

\section{Conclusions}
The present results provide a scheme allowing to control the typically highly irregular collective dynamics of polar active particles. 
In particular, we have seen that the bands occurring in the Vicsek model, which normally move in unpredictable
directions and irregularly merge and split up can be tamed
by applying a traveling wave-shaped potential, as can be realized e.g. using a micropatterned moving substrate or a
traversing optical lattice. 
We find that while most particles in the system self-organize into polarized bands which comove with the lattice at early times, they can later
experience a remarkable reversal, initiated by the counterpropagation of a single band which subsequently consumes all other 
bands in the system. The asymptotic state is a strongly polarized macroband which predictably moves opposite to the direction 
of the motion of the external substrate. 
This behaviour is representative in a large parameter window and can be controlled e.g. by tuning the relative speed of 
the active particles and the lattice. 
These results may inspire further research of the interface between nonlinear dynamics and active matter and 
perhaps also applications regarding collective targeted cargo delivery using polar active matter. 

\begin{acknowledgments}
A. Z. thanks  G. M. Koutentakis for fruitful discussions.
\end{acknowledgments}
%\newpage
\bibliographystyle{unsrt}

\bibliography{literature.bib}

\begin{thebibliography}{10}
\bibitem{footnote2}
Consider the dynamics of free, not fully overdamped particles $m
  \ddot{\vec{r}}_i+\gamma \dot{\vec{r}}_i=\gamma v_0 \vec{p}_i$ yielding the
  solution $\dot{\vec{r}}_i(t)=v_0
  \vec{p}_i\left(1-e^{-\frac{\gamma}{m}t}\right)$, i.e. the timescale on which
  a particle reaches the velocity $v_0 \vec{p}_i$ is $\approx
  \frac{m}{\gamma}$. We assume here that this time scale is much shorter than
  all other time scales in the system, including those applied by the
  travelling wave potential.
\bibitem{footnote1}
Choosing length and time units as $x_u=R, t_u=1/D_r$ shows that the parameter
  space has five essential dimensions: the Peclet number $v_0/(D_r R)$
  measuring the persistence length of the active particles (in bulk) in units
  of the particle radius, the reduced alignment rate $g/(\pi R^2 D_r)$,
  comparing the typical alignment rate of the particles with the Brownian decay
  of alignment, the reduced lattice depth $u_0/(\gamma R D_r)$ and the reduced
  frequency $\omega/D_r$ and wave vector $k R$ of the lattice (and the density
  $\rho R^2=2$). The mass unit is implicitly fixed as $m_u=D_r/\gamma$ yielding
  an energy unit of $E_u=m_u x_u^2/t_u^2=k T D_r R^2/D=3 kT/4$, where we have
  used the Einstein and the Stokes-Einstein-Debye relation.
\end{thebibliography}

\end{document}